\documentclass[twocolumn,aps,amssymb,amsmath,showpacs,preprintnumbers,superscriptaddress]{revtex4}

\input epsf
\usepackage[dvips]{color}
\begin{document}

\title{
Transverse-momentum dependent modification
 of dynamic texture in central Au+Au collisions at $\sqrt{s_{NN}}
 =200$ GeV
}

\affiliation{Argonne National Laboratory, Argonne, Illinois 60439}
\affiliation{University of Bern, 3012 Bern, Switzerland}
\affiliation{University of Birmingham, Birmingham, United Kingdom}
\affiliation{Brookhaven National Laboratory, Upton, New York 11973}
\affiliation{California Institute of Technology, Pasadena, California 91125}
\affiliation{University of California, Berkeley, California 94720}
\affiliation{University of California, Davis, California 95616}
\affiliation{University of California, Los Angeles, California 90095}
\affiliation{Carnegie Mellon University, Pittsburgh, Pennsylvania 15213}
\affiliation{Creighton University, Omaha, Nebraska 68178}
\affiliation{Nuclear Physics Institute AS CR, 250 68 \v{R}e\v{z}/Prague, Czech Republic}
\affiliation{Laboratory for High Energy (JINR), Dubna, Russia}
\affiliation{Particle Physics Laboratory (JINR), Dubna, Russia}
\affiliation{University of Frankfurt, Frankfurt, Germany}
\affiliation{Institute  of Physics, Bhubaneswar 751005, India}
\affiliation{Indian Institute of Technology, Mumbai, India}
\affiliation{Indiana University, Bloomington, Indiana 47408}
\affiliation{Institut de Recherches Subatomiques, Strasbourg, France}
\affiliation{University of Jammu, Jammu 180001, India}
\affiliation{Kent State University, Kent, Ohio 44242}
\affiliation{Lawrence Berkeley National Laboratory, Berkeley, California 94720}
\affiliation{Massachusetts Institute of Technology, Cambridge, MA 02139-4307}
\affiliation{Max-Planck-Institut f\"ur Physik, Munich, Germany}
\affiliation{Michigan State University, East Lansing, Michigan 48824}
\affiliation{Moscow Engineering Physics Institute, Moscow Russia}
\affiliation{City College of New York, New York City, New York 10031}
\affiliation{NIKHEF, Amsterdam, The Netherlands}
\affiliation{Ohio State University, Columbus, Ohio 43210}
\affiliation{Panjab University, Chandigarh 160014, India}
\affiliation{Pennsylvania State University, University Park, Pennsylvania 16802}
\affiliation{Institute of High Energy Physics, Protvino, Russia}
\affiliation{Purdue University, West Lafayette, Indiana 47907}
\affiliation{University of Rajasthan, Jaipur 302004, India}
\affiliation{Rice University, Houston, Texas 77251}
\affiliation{Universidade de Sao Paulo, Sao Paulo, Brazil}
\affiliation{University of Science \& Technology of China, Anhui 230027, China}
\affiliation{Shanghai Institute of Applied Physics, Shanghai 201800, China}
\affiliation{SUBATECH, Nantes, France}
\affiliation{Texas A\&M University, College Station, Texas 77843}
\affiliation{University of Texas, Austin, Texas 78712}
\affiliation{Tsinghua University, Beijing 100084, China}
\affiliation{Valparaiso University, Valparaiso, Indiana 46383}
\affiliation{Variable Energy Cyclotron Centre, Kolkata 700064, India}
\affiliation{Warsaw University of Technology, Warsaw, Poland}
\affiliation{University of Washington, Seattle, Washington 98195}
\affiliation{Wayne State University, Detroit, Michigan 48201}
\affiliation{Institute of Particle Physics, CCNU (HZNU), Wuhan 430079, China}
\affiliation{Yale University, New Haven, Connecticut 06520}
\affiliation{University of Zagreb, Zagreb, HR-10002, Croatia}

\author{J.~Adams}\affiliation{University of Birmingham, Birmingham, United Kingdom}
\author{M.M.~Aggarwal}\affiliation{Panjab University, Chandigarh 160014, India}
\author{Z.~Ahammed}\affiliation{Variable Energy Cyclotron Centre, Kolkata 700064, India}
\author{J.~Amonett}\affiliation{Kent State University, Kent, Ohio 44242}
\author{B.D.~Anderson}\affiliation{Kent State University, Kent, Ohio 44242}
\author{D.~Arkhipkin}\affiliation{Particle Physics Laboratory (JINR), Dubna, Russia}
\author{G.S.~Averichev}\affiliation{Laboratory for High Energy (JINR), Dubna, Russia}
\author{S.K.~Badyal}\affiliation{University of Jammu, Jammu 180001, India}
\author{Y.~Bai}\affiliation{NIKHEF, Amsterdam, The Netherlands}
\author{J.~Balewski}\affiliation{Indiana University, Bloomington, Indiana 47408}
% ...

\collaboration{STAR Collaboration}\noaffiliation

\date{\today}

\begin{abstract}
Correlations in the hadron distributions produced in relativistic Au+Au
collisions are studied in the discrete wavelet expansion method.  
The analysis is performed in the
space of pseudorapidity ($|\eta| \leq 1$) and azimuth (full $2\pi$)
in bins of transverse momentum ($p_t$) from $0.14 \leq p_t \leq 2.1$~GeV/$c$.
In peripheral Au+Au  collisions a correlation 
structure ascribed to minijet  fragmentation is observed.
It evolves with collision centrality and $p_t$ in a way not seen before which
suggests strong dissipation
of minijet fragmentation in the longitudinally-expanding medium.
\end{abstract}
\maketitle 
The study of the bulk properties of strongly interacting  matter under
extreme conditions at the Relativistic Heavy Ion Collider (RHIC)
is producing a number of tantalizing results~\cite{PhysicsToday}.
The physics of
central Au+Au collisions at RHIC is clearly much more complex than a mere
independent superposition  of nucleon-nucleon collisions,
while the issues of possible collectivity and of the 
degree of ``thermalization'' of the bulk hadronic medium remain open.  
Substantial equilibration, especially in a short-lived finite system,
may imply that during the evolution, there was a large number of degrees of
freedom involved, such as would occur in a partonic medium
~\cite{Feinberg} or quark-gluon plasma~\cite{Collins_Perry}.
Equilibration in heavy ion collisions
has been studied \emph{via} its effects on single particle  spectra and 
identified particle ratios.  
It progressively erases
correlations, starting with the smallest features
~\cite{Trainor,Shuryak-Stephanov}.  
Surviving correlations produced by hard scatterings
early in the collision
provide a sensitive monitor of the degree of equilibration of the medium.
In contrast, traversal of the QCD phase boundary
 may \emph{create}
specific dynamical correlations~\cite{QGP->correlations}.
Therefore correlations observed in the final state are potentially
affected by competing mechanisms. 
This makes the question of equilibration a quantitative one
and warrants a study of correlations among the majority of
hadrons over a range of momentum scales.
This Letter reports such a study.

In high energy elementary collisions,
hadrons originate from the fragmentation of a 
color-neutral system of partons.  In these systems
correlations are produced by local conservation of  
charge, flavor, energy and momentum in the strong interaction, and
by quantum statistics.
In high energy heavy ion collisions aspects of these elemental correlations
might persist, especially at high transverse momentum ($p_t$) since
the ``memory'' of the early hard partonic scattering is not easily erased there.  
In contrast,
minijets~\cite{Kajantie:1987pd} at lower $p_t$ 
are expected to have shorter mean free paths in the medium
and thus are more likely to dissipate,
erasing correlations.
The collision overlap density and size of
the interaction volume are changed by varying the centrality,
which might also control the degree of equilibration in these systems.
We study the correlation structure
in peripheral collisions, caused by minijets,
which evolves with centrality and $p_t$ in a manner suggesting
strong dissipation of minijet fragmentation
by the longitudinally expanding medium.

The data presented here were obtained with the
STAR Time Projection Chamber (TPC)~\cite{STAR_TPC}, 
mounted inside a solenoidal magnet.  Charged-particle tracking
with the TPC covers large acceptance well suited for precision
studies of correlation structures over a wide range of scales.
The minimum-bias event trigger discriminates on 
a neutral-spectator signal in the Zero
Degree Calorimeters~\cite{ZDC}.
Central events were selected by additionally requiring
a high charged particle multiplicity within $|\eta|<1$
in the Central Trigger Barrel scintillators~\cite{STAR_TPC}.
Accepted charged-particle tracks had $>15$ TPC
space points and $>$52\% of the estimated maximum
possible number of space points (to eliminate split tracks), 
passed within 3 cm of the event vertex
and were within the kinematic acceptance: 
$|\eta| \leq 1$, full $2\pi$ in  azimuth, and $0.14 \leq p_t \leq 2.1$~GeV/$c$.
Accepted events had their primary vertex within 25~cm of the geometric
center of the TPC longitudinally and had $\geq$15 accepted TPC tracks.
About 0.6~M central and 0.3~M peripheral events, recorded in the 
$\sqrt{s_{NN}}=200$ GeV run, were analyzed.

Two-point correlations and power spectra of 
point-to-point fluctuations are complementary measures used
to study the correlation structure of random fields (such as TPC events).
The former has computational complexity $O(N^2)$ ($N$ is event multiplicity).
The latter,
implemented via the discrete wavelet transform (DWT) method, is $O(N)$.
The DWT-based dynamic texture measure, defined below,
is used in this work and was
originally applied to relativistic 
Pb+Pb collisions by NA44~\cite{NA44}.

In this approach, the measured particle distribution $\rho(\phi,\eta)$ 
\emph{in a single event} is
expanded in the complete orthonormal wavelet basis of Haar~\cite{DWT}.
The scale of this basis is defined by the \emph{scaling 
function}
$g(x) = 1$ for $0\le x<1$ and 0 otherwise.
The function
$f(x) =     \{ 1 \mbox{\ for\ } 0\le x<0.5; 
                 -1 \mbox{\ for\ }  0.5 \le x<1;
                 \mbox{\ else}~ 0 
              \}$
is the wavelet function.
The experimental acceptance in $\eta$, $\phi$, and $p_t$ 
is split into equal bins in $\eta$,$\phi$ and $p_t$ bins exponentially
growing to equalize bin statistics.
To keep  notation simple but explicit, we introduce 
$\eta'\equiv (\eta+1)/2$ and $\phi'\equiv\phi/2\pi$
so that $\eta',\phi' \in [0,1]$.
The scaling function of the Haar basis in
two dimensions 
$G(\phi,\eta) = g(\phi')g(\eta')$
is just the bin acceptance (modulo units).
The wavelet functions $F^{\lambda}$ 
(the directional sensitivity mode $\lambda$ is either 
along azimuth $\phi$, pseudo-rapidity $\eta$, or diagonal $\phi\eta$
directions)
are
$F^{\phi\eta}=f(\phi')f(\eta')$,
$F^\phi=f(\phi')g(\eta')$,
$F^\eta=g(\phi')f(\eta')$.
We define a two dimensional  wavelet basis:
\begin{equation}
F^{\lambda}_{m,i,j}(\phi,\eta) =
 2^{m}F^{\lambda}(2^{m}\phi'-i,2^{m}\eta'-j),
\label{wavelet_2D}
\end{equation}
where $m \geq 0$ is 
the integer
scale fineness index~\cite{ref15},
integers $i$ and $j$ index the positions of bin centers in 
$\phi'$ and $\eta'$, and $0 \le i,j < 2^m$.
Scaling functions $G_{m,i,j}(\phi,\eta)$ are constructed analogous
to Eq.\ref{wavelet_2D}.  Arbitrary density $\rho(\phi,\eta)$ is expanded as
\begin{equation}
\label{eq:rho}
\rho(\phi,\eta) = \langle \rho, G_{0,0,0}\rangle G_{0,0,0}
 + \sum_{m,i,j,\lambda} 
\langle \rho, F^{\lambda}_{m,i,j} \rangle F^{\lambda}_{m,i,j},
\end{equation}
where $\langle \rho, G \rangle$ and $\langle \rho, F^{\lambda} \rangle$
are expansion coefficients obtained by projecting density $\rho(\phi,\eta)$ onto
the basis functions.


In practice $m \le m_{\textrm{max}}$, where $m_{\textrm{max}}$ is
the finest scale limited by
track resolution and,
due to the needs of event mixing, by the number of available events.
The coarser scales correspond to successively re-binning the track 
distribution.
The analysis is best visualized by considering the scaling function
$G_{m,i,j}(\phi,\eta)$ as binning the track distribution 
$\rho(\phi,\eta)$
in bins $i$,$j$
of given fineness $m$, while
the wavelet expansion
coefficients $\langle \rho, F^{\lambda}_{m,i,j}\rangle$
give the difference distribution for data with binning
one step finer.
The wavelet expansion coefficients were calculated using the code
{\sc  waili}~\cite{WAILI}.

The \emph{power spectrum} is defined as
\begin{equation}
P^\lambda(m) =
2^{-2m}
\overline{\sum_{i,j}\langle \rho,F^\lambda_{m,i,j}\rangle^2} ,
\label{eq:P_m}
\end{equation}
where the overline denotes an average over events.
$P^\lambda(m)$ is independent of $m$ for an uncorrelated $\rho$.
However, for physical events  $P^\lambda$  depends
on $m$ due to the presence of \emph{static texture} features such as
acceptance asymmetries and imperfections (albeit minor in STAR),
and non-uniformity of $\,dN/\,d\eta$.
To remove these known features from the analysis a reference is
constructed from mixed events starting with individual ($\phi,\eta$)
pixels of true events at the finest scale used in the analysis ($16\times16$).
A ``mixed event'' 
consists of $16\times16$ ($\phi,\eta$)
pixels from true events, where each pixel is taken from different,
but similar, real events.  
The power spectrum $P^\lambda_{\textrm{mix}}$ is obtained from Eq.~\ref{eq:P_m} using
the expansion coefficients in Eq.~\ref{eq:rho}. 
$P^\lambda_{\textrm{mix}}$ contains static, experimental track density artifacts
plus statistical noise.
The quantity of interest is the
difference, $P^\lambda_{\textrm{dyn}} \equiv P^\lambda_{\textrm{true}} - P^\lambda_{\textrm{mix}}$, 
called
\emph{dynamic texture}~\cite{NA44}.

In studying the dynamic texture data as a function of $p_t$, the desirable 
normalization is such that the results are 
independent of $p_t$ bin size under the assumption of large-scale
correlations in $p_t$ ({\it i.e.} larger than the $p_t$ acceptance).
In this case for increasing number of particles $N$ in an increasing $p_t$ bin,
$P^\lambda_{\textrm{dyn}} \propto N^2$
while $P^\lambda_{\textrm{mix}} \propto N$, 
being a Poissonian variance.
Therefore we present the data as the combined quantity
$P^\lambda_{\textrm{dyn}}/P^\lambda_{\textrm{mix}}/N$.

 Systematic error can be introduced in
$P^\lambda_{\textrm{dyn}}$ by the process of event mixing.  For example,
events with different vertex positions along the beam axis
are reconstructed with slightly different efficiencies and
acceptances with respect to $\eta$.
This variable efficiency may fake a dynamic texture effect in $\eta$.
In order to minimize such errors,
events were grouped into event classes with similar
multiplicity (within 50) and vertex position (within 10~cm).
$P^\lambda_{\textrm{dyn}}$ was constructed using only events within each
of these two classes.  Results showed no vertex 
dependence.
The upper limit on the systematic error due to $z$-vertex position
variation is set by the
statistical error of the data, shown in the figures.

Event centrality in this analysis is characterized by the accepted number
of quality tracks in the TPC and expressed as a percentage of the total inelastic
cross-section, as before\cite{STAR_centrality}.
Event classes in multiplicity are grouped to form two centrality classes:
\emph{central}, with 4\% of the most central events, and \emph{peripheral}, 
with event centrality
varying between 60\% and 84\%.
The  {\sc hijing} ~\cite{HIJING_model} generator events for the Monte Carlo comparison
are selected to match these centrality ranges.

Track splitting (one particle reconstructed as $>1$ track)
contributions were eliminated by track quality requirements.
Track merging ($>1$ particles reconstructed as one track)
mocks up anticorrelations and can induce systematic error.
To estimate this effect, central {\sc hijing} events were filtered
with an algorithm emulating track recognition properties
of the TPC~\cite{ref16}.
The simulation results can be expressed as a set of
coefficients relating $P_{\textrm{true}}$,
$P_{\textrm{mix}}$ and $\,dN/\,dp_t$ in the original and filtered {\sc hijing} data.
An estimate of track  merging effects in the data was obtained from the inverse
of these coefficients.
The resulting systematic error
was estimated to be $0.5\times10^{-4}$.
Systematic error due to non-primary background  was estimated
assuming that the correlations between true primary and non-primary particles
could be anything from zero to that 
of primary particles themselves.
The systematic error estimate was taken to be half the difference between
these two limits which is 10\% of the signal at 
$p_t=0.2$~GeV/$c$, falling
to 3.5\% at $p_t=1$~GeV/$c$.
This estimate applies to both centrality classes.

\begin{figure}
\epsfxsize=8.6cm
\epsfbox{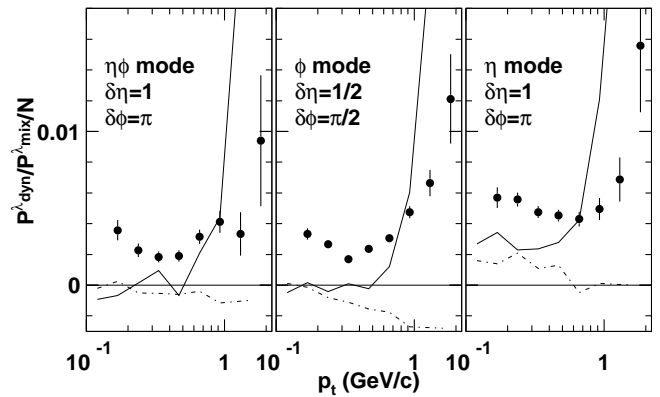}
\caption{
Peripheral events: normalized dynamic texture for fineness scales $m=0,1,0$
from left to right panels, respectively,
as a function of $p_t$.
\textcolor{black}{$\bullet$} -- STAR data;
solid line -- {\sc hijing} without jet quenching;
dash-dotted line -- {\sc hijing} without jets.
}
\label{STAR_peripheral}
\end{figure}

\begin{figure}[htb]
\epsfxsize=8.6cm
\epsfbox{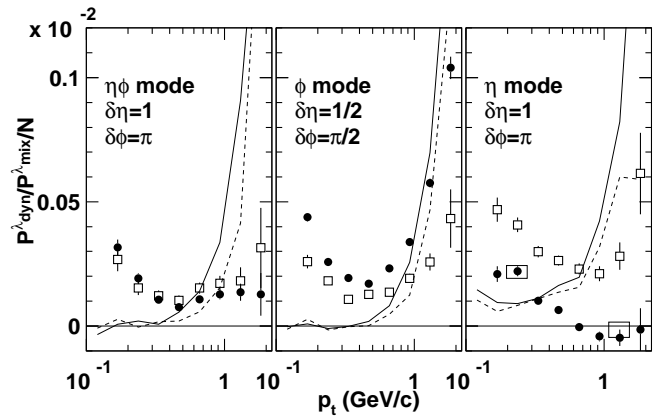}
\caption{
Central events:
normalized dynamic texture for fineness scales $m=0,1,0$
from left to right panels, respectively,
as a function of $p_t$.
\textcolor{black}{$\bullet$} -- STAR data;
solid line -- {\sc hijing} without jet quenching;
dashed line --  {\sc hijing} with quenching;
\textcolor{black}{$\Box$} -- peripheral
STAR data from Fig.~\ref{STAR_peripheral}
renormalized as described in the text.
The rectangles around two chosen points show the estimated systematic errors.
}
\label{STAR_central}
\end{figure}

Fig.~\ref{STAR_peripheral} presents  measured  large-scale 
dynamic texture in peripheral 
 collisions
compared with {\sc hijing} predictions where fineness scale
$m=0$ is used
for the $\eta\phi$ and $\eta$ modes and fineness $m=1$
 is used for the $\phi$ mode.
The finer scale for $\delta\phi$ is used so that the angular
coverages subtended by $\delta\phi$ for $m=1$ and $\delta\eta$ for $m=0$ are comparable.
This scale, with $\phi$-bin size $\delta\phi = \pi/2$~\cite{ref15},
is dominated by elliptic flow~\cite{v2_RHIC}.
The  {\sc hijing} calculations without jet quenching show
a region of approximately constant signal near $p_t \sim 0.5~GeV/c$
followed by an increase for $p_t > 0.8$ GeV/$c$,
obtained by ``turning on'' jets in the model.
In that $p_t$ range the STAR data  also increase with $p_t$.
Momentum conservation suppresses the difference 
in the numbers of tracks emitted in the opposite directions.
This effect is absent in  the mixed events, resulting
in negative $P^\lambda_{\textrm{dyn}}$, seen in $\phi$ when jets in {\sc hijing} are ``off''.
Comparing the two simulations in Fig.~\ref{STAR_peripheral} we conclude that fluctuations in
local hadron density due to jets are observable in peripheral RHIC collisions
at $0.8<p_t<2$ GeV/$c$.
This supports but does not prove the identification of similar signals 
in the data at these $p_t$ with minijets.
Without ruling out other sources of angular correlations at such $p_t$,
we use Occam's razor to adopt the well established effect -- 
fragmentation of semi-hard scattering products 
(jets or minijets) -- as the explanation.

Central event data and {\sc hijing} predictions with 
and without  jet quenching are shown in Fig. \ref{STAR_central}.
The most striking difference here compared to the peripheral data in
Fig.~\ref{STAR_peripheral} is the reduction in the magnitude of the
$P^\eta_{\textrm{dyn}}$ at larger $p_t > 0.6$~GeV/$c$, the data becoming slightly
negative near 1~GeV/$c$ in sharp contrast to the jet-like behavior
predicted by {\sc hijing}.
The perturbative partonic energy loss model of jet quenching in {\sc hijing} seems to miss
the correlation aspect of the picture, at least at these $p_t$.
In the absence of a successful theory to describe the effect,  
we formulate and test a ``null hypothesis'': the correlation structure 
$P^\lambda_{\textrm{dyn}}/P^\lambda_{\textrm{mix}}$ in Au+Au collisions is independent of
centrality.
Then, the difference in 
$P^\lambda_{\textrm{dyn}}/P^\lambda_{\textrm{mix}}/N$ in central and peripheral events
(including the $p_t$ trends) is due to the difference in $1/N$ 
({\it i.e.} in $\,dN/\,dp_t$, for $N\equiv N(p_t) = \int_{p_t \textrm{bin}} \,dN(p_t)$)
~\cite{ref17}.
Shown in Fig.~\ref{STAR_central} as symbol $\Box$ are the peripheral data 
from Fig.~\ref{STAR_peripheral}, rescaled under an assumption of the ``null hypothesis'' by
$\times N(p_t)|_{\textrm{periph}}/N(p_t)|_{\textrm{centr}}$.
The left panel shows that  the $\eta\phi$-mode
is less affected by centrality,
reflecting a superposition of the opposite centrality trends in $\eta$ and $\phi$.
We hypothesize that the deviation of the STAR data from the ``null hypothesis''
in $\eta$ in the otherwise correlated system
points to a \emph{randomization} (dissipation) of minijet structure in the 
longitudinal direction.
Longitudinal expansion of the hot, dense medium formed early in the collision 
singles out the $\eta$ direction and is
likely to be part of the dissipation mechanism.
If so, at $p_t>0.6$~GeV/$c$ we may be observing an effect of the longitudinally 
expanding medium on parton fragmentation or hadronization.

In each panel of Figs.~\ref{STAR_peripheral} and \ref{STAR_central} the
dynamic texture data increase with decreasing $p_t$ for $p_t<0.4$ GeV/$c$.
Data stay non-zero at low $p_t$ for all three modes in the experiment and for the
$\eta$-mode in {\sc hijing}.  In this $p_t$ range, the correlations are
likely dominated by centrality-dependent effects such as the final
state quantum statistical intensity interference, Coulomb effect and
longitudinal string fragmentation physics, simulated in {\sc hijing}.
Modification of the latter effect with centrality is the subject of a
separate publication\cite{STAR_CD}.

\begin{figure}
\epsfxsize=8.6cm
\epsfbox{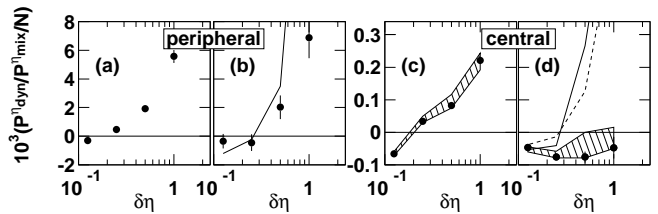}
\caption{Scale dependence of the dynamic texture in
peripheral and central events.(a,c): $0.2<p_t<0.28$,
(b,d): $1.1<p_t<1.5$  GeV/$c$.
\textcolor{black}{$\bullet$} -- STAR data;
solid line -- {\sc hijing} without jet quenching;
dashed line -- {\sc hijing} with quenching.
A systematic error estimate
is shown as a hatched area.
Errors on different scales are estimated independently.
}
\label{scale_dependence}
\end{figure}

Fig.~\ref{scale_dependence} 
shows a scale dependence of the $\eta$-mode
in the low and higher $p_t$ intervals.
At low $p_t$, the peripheral and central trends qualitatively agree, whereas at
higher $p_t$, a modification with centrality is seen, which testifies to the presence
of new physics at higher $p_t$.
The reduction of the dynamic texture in central events
with respect to both {\sc hijing}  and the peripheral STAR data
is most dramatic at the coarser scales.
The longitudinal expansion correlates  $\eta$ with the longitudinal coordinate $z$, 
and $z$ -- with time. 
Final state particles with large $\delta\eta$ are more likely
to be separated by a space-like interval. 
Thus, the larger  $\delta\eta$ correlations are more likely to have their 
cause in the particles' common past,
reflecting the early stage of the system, 
whereas the fine scale features are formed later under conditions little different
from peripheral collisions or conventional hadronic models.
The negative $P^{\eta}_{\textrm{dyn}}$ in Fig.~\ref{scale_dependence}(d) points
to the presence of an anticorrelation mechanism, which could include
existence of a characteristic scale in the longitudinal separation of
hadrons in the course of hadronization.  
Lack of scale dependence in Fig.~\ref{scale_dependence}(d), relative to
Fig.~\ref{scale_dependence}(b), may be contrasted with progressive reduction of
small-scale Fourier harmonics from hadronic diffusion discussed
in~\cite{Shuryak-Stephanov}. Alternatively, pre-hadronic transport on $\eta$
involving partonic diffusion could provide a more efficient
equilibration mechanism. Other mechanisms such as convective turbulent
transport~\cite{equilibration_mechanisms} might also play a role.
The reduction of dynamic
texture reported in this Letter provides a new quantitative argument
in favor of equilibration or dissipation effects.  
However, we observe that the hadronic final state is not correlation free,
even for central events.

In summary, a non-trivial picture emerges when the DWT power spectrum
technique is applied for the first time to Au+Au collision data from
RHIC.  Large-scale ($\delta\eta =1$) angular correlations for
$p_t<2.1$~GeV/$c$ are observed in peripheral events and identified
with minijets.  In central events, those correlations are suppressed
with increasing $p_t$ and $\delta\eta$.  This indicates a major change
in the properties of the medium with increasing collision centrality,
implying the development of a dissipative medium.  In the course of
its longitudinal expansion, this hypothetic medium influences via
interactions the structure of correlations, inherited from the
kinematics of the initial-state semi-hard scattering, causing their
dissipation and partial equilibration.

We thank the RHIC Operations Group and RCF at BNL, and the
NERSC Center at LBNL for their support. This work was supported
in part by the HENP Divisions of the Office of Science of the U.S.
DOE; the U.S. NSF; the BMBF of Germany; IN2P3, RA, RPL, and
EMN of France; EPSRC of the United Kingdom; FAPESP of Brazil;
the Russian Ministry of Science and Technology; the Ministry of
Education and the NNSFC of China; Grant Agency of the Czech Republic,
FOM and UU of the Netherlands, DAE, DST, and CSIR of the Government
of India; Swiss NSF; and the Polish State Committee for Scientific 
Research.


\begin{thebibliography}
\bibitem{ref15}
We specify scale by the $m$ index or by the
bin size $\delta x$ in the space of kinematic variable $x$.
For acceptance $\Delta x$, scale (bin size)
$\delta x = \Delta x/2^{m+1}$, $m \geq 0$.
\bibitem{ref16}
The algorithm merges close tracks, based on known
close track recognition efficiencies in STAR, where all
two-track separation distances are evaluated
at three radii in the TPC; close pairs are merged.
\bibitem{ref17}
This hypothesis is not identical with assuming that A+A collisions are a 
superposition of independent $NN$ collisions -- the latter assumption
excludes modifications of $p_t$ spectra with centrality, which we allow.
\end{thebibliography}
\end{document}